\documentclass[citeauthoryear,11pt]{article}
\usepackage{epsfig}
\begin{document}
\input amssym.def 
\input amssym
\hfuzz=5.0pt
%
%
%
%
\def\vec#1{\mathchoice{\mbox{\boldmath$\displaystyle\bf#1$}}
{\mbox{\boldmath$\textstyle\bf#1$}}
{\mbox{\boldmath$\scriptstyle\bf#1$}}
{\mbox{\boldmath$\scriptscriptstyle\bf#1$}}}
\def\mbf#1{{\mathchoice {\hbox{$\rm\textstyle #1$}}
{\hbox{$\rm\textstyle #1$}} {\hbox{$\rm\scriptstyle #1$}}
{\hbox{$\rm\scriptscriptstyle #1$}}}}
\def\operatorname#1{{\mathchoice{\rm #1}{\rm #1}{\rm #1}{\rm #1}}}
\chardef\ii="10
\def\widehat{\mathaccent"0362 }
\def\widetilde{\mathaccent"0365 }
\def\vphi{\varphi}
\def\vrho{\varrho}
\def\vtheta{\vartheta}
\def\ih{{\i\over\hbar}}
\def\CD{{\cal D}}
\def\CH{{\cal H}}
\def\CL{{\cal L}}
\def\CP{{\cal P}}
\def\CV{{\cal V}}
\def\half{{1\over2}}
\def\bhalf{\hbox{$\half$}}
\def\viert{{1\over4}}
\def\bviert{\hbox{$\viert$}}
\def\dfrac#1#2{\frac{\displaystyle #1}{\displaystyle #2}}
\def\pathint#1{\int\limits_{#1(0)=#1'}^{#1(T)=#1''}\CD #1(t)}
\def\hbarm{{\dfrac{\hbar^2}{2m}}}
\def\pmb#1{\setbox0=\hbox{#1}
    \kern-.025em\copy0\kern-\wd0
    \kern.05em\copy0\kern-\wd0
    \kern-.025em\raise.0433em\box0}
\def\bbbm{{\rm I\!M}}
\def\bbbn{{\rm I\!N}}                                
\def\bbbr{{\rm I\!R}}                                
\def\bbbz{{\mathchoice {\hbox{$\sf\textstyle Z\kern-0.4em Z$}}
{\hbox{$\sf\textstyle Z\kern-0.4em Z$}}
{\hbox{$\sf\scriptstyle Z\kern-0.3em Z$}}
{\hbox{$\sf\scriptscriptstyle Z\kern-0.2em Z$}}}}    
\def\Cl{\operatorname{Cl}} 
\def\Coulomb{\operatorname{Coulomb}} 
\def\Higgs{\operatorname{Higgs}} 
\def\max{\operatorname{max}} 
\def\PSL{\operatorname{PSL}} 
\def\dt{\d t}
\def\d{\operatorname{d}}
\def\e{\operatorname{e}}
\def\i{\operatorname{i}}
 
\begin{titlepage}
\centerline{\normalsize DESY 98--112 \hfill ISSN 0418 - 9833}
\centerline{\normalsize August 1998\hfill}
\centerline{\normalsize quant-ph/9808060\hfill}
\vskip.3in
\message{TITLE:}
\message{ON THE PATH INTEGRAL TREATMENT FOR AN AHARONOV--BOHM FIELD ON THE 
HYPERBOLIC PLANE}
\begin{center}
{\Large ON THE PATH INTEGRAL TREATMENT FOR AN 
\vskip.05in
AHARONOV--BOHM FIELD ON THE HYPERBOLIC PLANE}
\end{center}
\vskip.3in
\begin{center}
{\Large Christian Grosche}
\vskip.2in
{\normalsize\em II.\,Institut f\"ur Theoretische Physik}
\vskip.05in
{\normalsize\em Universit\"at Hamburg, Luruper Chaussee 149}
\vskip.05in
{\normalsize\em 22761 Hamburg, Germany}
\end{center}
\vfill
\begin{center}
{ABSTRACT}
\end{center}
\smallskip
\noindent
{\small
In this paper I discuss by means of path integrals the quantum dynamics of a 
charged particle on the hyperbolic plane under the influence of an Aharonov--Bohm 
gauge field. The path integral can be solved in terms of an expansion of the 
homotopy classes of paths. I discuss the interference pattern of scattering by 
an Aharonov--Bohm gauge field in the flat space limit, yielding a characteristic 
oscillating behavior in terms of the field strength. In addition, the cases of the 
isotropic Higgs-oscillator and the Kepler--Coulomb potential on the hyperbolic 
plane are shortly sketched. 
}

\end{titlepage}

%
 
\normalsize
\section{Introduction}
\message{Introduction}
The Aharonov--Bohm gauge field has a long history, beginning in 1959 by a
classical paper by Aharonov and Bohm \cite{AB}. The effect has been well 
studied and well confirmed \cite{Book:A-B}, but not necessarily well
understood. It describes the motion of charged particles, i.e.~electrons, which 
are scattered by an infinitesimal thin solenoid. The magnetic vector potential 
$\vec A$ of the solenoid produces a magnetic field which is essentially 
$\delta$-like, i.e., its support is an infinitesimal thin
solenoid, and it is vanishing everywhere else. Geometrically this experimental
set-up corresponds to the quantum motion of a particle (which we consider as
spin-less) in $\bbbr^2$, where a point has been removed with the consequence 
that topologically $\bbbr^2$ becomes no longer connected. Since the solenoid 
is assumed impenetrable, the space of the particle motion $\bbbm$ is the
Euclidean plane minus the cross section of the solenoid. Everywhere in 
$\bbbm$, $\vec\nabla\times\vec A=0$ and hence $\vec A=\vec\nabla f(r)$,
where $f(r)$ is an arbitrary scalar function of $r=|\vec x|,\vec x\in\bbbr^2$. 
Classically, a charged particle is not affected at all by the solenoid.
However, in quantum mechanics, the particle's wave function picks up in a 
scattering experiment a phase factor according to
\begin{equation}
  \Psi_\alpha(\vec x)=\Psi_0(\vec x)\exp\Bigg(\dfrac{\i e}{\hbar c}
       \int_{\hbox{\small path $\alpha$}}\vec A\cdot\d\vec x\Bigg)\enspace,
\end{equation}
where $ \Psi_0(\vec x)$ is the vector potential-free solution. The wave-function
$\Psi$ effective to a measurement is the sum of solutions corresponding
to inequivalent paths, i.e., $\Psi=\sum_\alpha\Psi_\alpha$. Topologically the
paths $\alpha$ can be distinguished by their winding numbers $n$, thus giving
rise to infinitely many homotopy classes designated by the number $n$.

Path integral treatments of the Aharonov--Bohm effect in the Euclidean plane are 
due to Bernido and Inomata \cite{BEIN}, Gerry and Singh \cite{GSa}, Liang 
\cite{LIANG}, and Schulman \cite{SCHUHc}. Harmonic interactions have been dealt 
with in \cite{KICA}, the Coulomb--Kepler potential have been taken into 
account by, e.g.\ \cite{CGHC,DRCAKI,HHKR}, \linebreak \cite{KNg,LINDHc,PARKb}, 
relativistic particles by, e.g.\ \linebreak \cite{BERNe,GARI,HHKR,LEVAN}, 
\linebreak \cite{LINDHc,PARKb}, 
and a more comprehensive bibliography can be found in, e.g.~\cite{Book:A-B,GRSh}.

Path integrals,\,e.g.\,\cite{FH,GROad,GRSh}, \linebreak \cite{KLEo}, and 
\cite{SCHUHd} provide us with global information of the quantum motion, 
including the topological effects on the wave-function. If we want to study the 
Aharonov--Bohm effect by means of path integrals \cite{BEIN,GSa,LIANG} we consider 
the time evolution from $t=0$ to $t=T$ of the wave-function of a particle 
according to
\begin{equation}
\Psi_\alpha(\vec x'';T)=\sum_\beta\int K_{\alpha\beta}(\vec x'',\vec x';T)
 \Psi_\alpha(\vec x';0)\,\d\vec x'\enspace,
\end{equation}
where
\begin{equation}
K_{\alpha\beta}(\vec x'',\vec x';T)= K_0(\vec x'',\vec x';T)
\exp\Bigg[\dfrac{\i e}{\hbar c}\Bigg(\int_{\hbox{\small path $\alpha$}}^{\vec x''}- 
       \int_{\hbox{\small path $\beta$}} ^{\vec x'}\Bigg)\vec A\cdot\d\vec x\Bigg]
\enspace,
\end{equation}
and this leads us the the formal expression separating the sum over
$\alpha$ and $\beta$ (under the assumption the separation is well-defined)
\begin{equation}
  \sum_{\alpha,\beta}K_{\alpha\beta}\Psi_\beta=K\sum_\beta\Psi_\beta\enspace.
\end{equation}
Provided the paths $\alpha,\beta$ cover in an idealized experiment the whole range
from minus infinity to plus infinity, we can express the separation of the time
evolution of the particle according to
\begin{equation}
  K(\vec x'',\vec x';T)=\sum_{n=-\infty}^\infty K_n(\vec x'',\vec x';T)\enspace,
\label{KvecxT}
\end{equation} 
where $n=0$ denotes the unperturbed case in $\bbbr^2$, i.e., we obtain the free 
propagator on the entire $\bbbr^2$. For the final result we obtain for the Feynman 
kernel the following form, e.g.\ \cite{BEIN,GRSh,LIANG}
\begin{equation}
  K(\vec x'',\vec x';T)=\dfrac{m}{2\pi\i\hbar T}
  \exp\bigg(\dfrac{\i m}{2\hbar T}({r'}^2+{r''}^2)\bigg)
  \sum_{n=-\infty}^\infty\e^{\i n(\vphi''-\vphi')}
  I_{|n-\xi|} \bigg(\dfrac{mr'r''}{\i\hbar T}\bigg)\enspace.
\end{equation} 
Here, two-dimensional polar coordinates $(r,\vphi)$ have been used, and $\xi=
e\Phi/2\pi\hbar c$ with $\Phi=B\times\hbox{area}$ the magnetic flux.

\section{Aharonov--Bohm Field on the Hyperbolic Plane}
\message{Aharonov--Bohm Field on the Hyperbolic Plane}
In this paper I would like to give a path integral treatment of the
Aharonov--Bohm effect on the hyperbolic plane \cite{KURORU}, i.e., the
scattering of (spin-less) electrons by an Aharonov--Bohm field on leaky tori.
Such systems play an important r\^ole in the theory of quantum chaos, e.g.\
\cite{GUTc}. The hyperbolic plane, respectively Lobachevsky space, is defined as 
one sheet of the double sheeted hyperboloid
\begin{equation}
  \vec u^2=u_0^2-u_1^2-u_2^2=R^2\enspace,\qquad u_0>0\enspace.
\end{equation}
The model of the upper-half plane $U=\{\Im(z)=y>0|z=x+\i y\}$ endowed with the 
metric has the form (where I have set for simplicity $R=1$)
\begin{equation}
  \d s^2=\dfrac{\d x^2+\d y^2}{y^2},\enspace,\qquad x\in\bbbr,y>0\enspace.
\end{equation}
Alternatively I can also consider the unit disc model $D=\{z=r\,\e^{\i\vtheta}|
r<1,\vtheta\in[0,2\pi)\}$)
\begin{equation}
 \d s^2=4\dfrac{\d r^2+r^2\d\vtheta^2}{(1-r^2)^2}\enspace,\qquad
 r<1,\vtheta\in[0,2\pi)\enspace,
\end{equation}
and the pseudosphere $\Lambda=\{z=\i\tanh(\tau/2)\,\e^{-\i\vphi}|
\tau>0,\vphi\in[0,2\pi)\}$
\begin{equation}
 \d s^2=\d\tau^2+\sinh^2\tau\,\d\vphi^2\enspace,\qquad\tau>0,\vphi\in[0,2\pi)
  \enspace.
\end{equation}
$U$, $D$ and $\Lambda$ are three coordinate space representations out of nine 
of the hyperbolic plane \cite{GROPOc,GROad,OLE}. Plane waves have the asymptotic
representation $\propto y^{1/2\pm\i k}$ (e.g.~on $U$, $k$ the wave-number), 
$\e^{-(\pm\i k+1/2)\tau}$ (on $\Lambda$), and the
coordinate origin is $r=0$ (on $D$), $\tau=0$ (on $\Lambda$), and $z=\i$ (on $U$),
respectively. The isometries on the hyperbolic plane are M\"obius transformations
corresponding to the symmetry group $\PSL(2,\bbbr)$, and magnetic fields give rise 
to the consideration of automorphic forms in the
theory of the Selberg trace formula \cite{HEJb}.

Constant magnetic fields on the hyperbolic plane have been studied in, e.g.\ 
\cite{COM,FAY,PNUELI}, and by means of path integrals in \cite{GROb,GROd}. 
The path integral formulation for a particle on the hyperbolic plane subject 
to a constant magnetic field on $\Lambda$ has the form \cite{GROd}
(I implicitly assume that the constant negative curvature of the hyperbolic plane,
i.e., the two-dimensional hyperboloid equals one, $\vec u\in\Lambda$)
\begin{eqnarray}  & &\!\!\!\!\!\!\!\! 
  K(\vec u'',\vec u';T)\equiv K(\tau'',\tau',\vphi'',\vphi';T)
    \nonumber\\   & &\!\!\!\!\!\!\!\! 
  =\pathint{\tau}\sinh\tau\pathint{\vphi}
    \nonumber\\   & &\!\!\!\!\!\!\!\! \qquad\times
  \exp\Bigg\{\ih\int_0^T\bigg[{m\over2}(\dot\tau^2+\sinh^2\tau\dot\vphi^2)
  -b(\cosh\tau-1)\dot\vphi
  -\dfrac{\hbar^2}{8m}\bigg(1-\dfrac{1}{\sinh^2\tau}\bigg)\bigg]\d t\Bigg\}
    \nonumber\\   & &\!\!\!\!\!\!\!\!
 =\exp\bigg(-{\i\hbar T\over8m}\bigg)\lim_{N\to\infty}
 \bigg(\dfrac{m}{2\pi\i\hbar\epsilon}\bigg)^N\prod_{j=1}^{N-1}
  \int_0^\infty\sinh\tau_j\,\d\tau_j\int_0^{2\pi}\d\vphi_j
    \nonumber\\   & &\!\!\!\!\!\!\!\!\qquad\times
  \exp\left[\ih\sum_{j=1}^N\bigg({m\over2\epsilon}
  \Big(\Delta^2\tau_j+\widehat{\sinh^2\tau_j}\Delta^2\vphi_j\Big)
  -b(\widehat{\cosh\tau_j-1})\Delta\vphi_j-{\epsilon\hbar^2\over8m\sinh^2\tau_j}
  \bigg)\right]
    \nonumber\\   & &\!\!\!\!\!\!\!\!
  =\sum_{l=-\infty}^\infty\left[
    \sum_{N=0}^{N_{\max}} \e^{-\i E_nT/\hbar}
    \Psi^b_{Nl}(\tau'',\vphi'') \Psi_{Nl}^{b\,*}(\tau',\vphi')
    +\int_0^\infty\!\!\d k\,\e^{-\i E_kT/\hbar}
    \Psi^b_{kl}(\tau'',\vphi'') \Psi_{kl}^{b\,*}(\tau',\vphi')\right]\,.
    \nonumber\\   & &
\label{Kpath-Bfeld}
\end{eqnarray}
Here $b=eB/\hbar c$, with $B$ the strength of the magnetic field, $c$ denotes the
velocity of light. For the magnetic field $\vec B$ I have chosen the gauge
\begin{equation}
\vec A=\bigg(\begin{array}{c}A_\tau\\ A_\vphi\end{array}\bigg)=
    B(\cosh\tau-1)\bigg(\begin{array}{c} 0\\ 1\end{array}\bigg)\enspace.
\label{Atau}
\end{equation}
Due to $\d B=(\partial_\tau A_\vphi-\partial_\vphi A_\tau)\,\d\tau\wedge\d\vphi=
(m/2)B\sinh\tau \d\tau\wedge\d\vphi$, $\d B$ has the form {\it constant $\times$ 
volume form\/} and can thus interpreted indeed as a constant magnetic field. In the 
lattice formulation I have taken \cite{GROad,GRSh} $\Delta q_j=q_j-q_{j-1}$, 
$q_j=q(t_j)$, $t_j=j\epsilon$, $j=1,\dots,N$, $\epsilon=T/N$, $N\to\infty$, 
$\widehat{f^2(q_j)}\equiv f(q_{j-1})f(q_j)$, for any function $f$ of the coordinates.
The bound state solutions are given by
\begin{eqnarray} 
  \Psi^b_{N,l}(\tau,\vphi)&=&
  \bigg[{N!(2b+|l|)\Gamma(2b-N+|l|)\over4\pi(N+|l|)!\Gamma(2b-N)}\bigg]^\half
     \nonumber\\   & &\times
  \e^{\i l\vphi}\bigg(\tanh{\tau\over2}\bigg)^{|l|}
  \bigg(1-\tanh^2{\tau\over2}\bigg)^{b-N}
  P_N^{(|l|,2b-2N-1)}\bigg(1-2\tanh^2{\tau\over2}\bigg)\enspace,\qquad
              \\  
  E_N&=&{\hbar^2\over2m}\Bigg[b^2+{1\over4}-\bigg(b-N-\half\bigg)^2\Bigg]\enspace,
       \qquad(N=0,1,\dots\leq N_{\max}<b-\bhalf)\enspace.\qquad
\label{uplaneEN}
\end{eqnarray}
$P_n^{(a,b)}(x)$ are Jacobi polynomials \cite{GRA}. The energy-levels (\ref{uplaneEN}) 
are the Landau levels on the hyperbolic plane. This is in complete analogy to the 
flat space case, where the Landau levels are $E_n=\hbar\omega(n+\half)$ with $\omega
=eB/\hbar c$ the cyclotron frequency, and the bound states are described by Laguerre 
polynomials, e.g.~\cite{GRSh}. The flat space limit can be recovered \cite{GROPOc} 
by re-introducing the constant curvature $k=1/R$ $(R>0)$, redefining $E_N\to E_N/R^2,
b\to bR^2$ (note $b(\cosh\tau-1)\to br^2R^2/2$, $r>0$ the polar variable in $\bbbr^2$, 
as $R\to\infty$), and considering the limit $R\to\infty$.

For the continuous states the wave-functions and the energy spectrum, 
respectively, I obtain
\begin{eqnarray} 
  \Psi_{k,l}^b(\tau,\vphi)&=&{1\over\pi|l|!}
  \sqrt{k\sinh2\pi k\over4\pi}\,
  \Gamma\bigg({1+\i k\over2}+b+|l|\bigg)\Gamma\bigg({1+\i k\over2}-b\bigg)
     \nonumber\\   & &\qquad\times
  \e^{\i l\vphi}\bigg(\tanh{\tau\over2}\bigg)^{|l|}
            \bigg(1-\tanh^2{\tau\over2}\bigg)^{\half+\i k}
     \nonumber\\   & &\qquad\times
  {_2}F_1\bigg(\half-\i k+b+|l|,\half+\i k-b;1+|l|;
               \tanh^2{\tau\over2}\bigg)\enspace,\qquad
              \\  
  E_k&=&{\hbar^2\over2m}\bigg(k^2+b^2+{1\over4}\bigg)\enspace.
\end{eqnarray}
$_2F_1(a,b;c;z)$ is the hypergeometric function, and $k>0$ denotes the wave-number. 
I note that a minimum strength of $B$
is required in order that bound states can occur, and only a finite
number of bound states can exist. For the case that the magnetic field vanishes 
I obtain \cite{GRSc} (e.g.~\cite{GRA} for the relation of the Legendre functions 
to the hypergeometric function)
\begin{eqnarray}
 \Psi_{k,l}&=&\sqrt{k\sinh\pi k\over2\pi^2}\,\Gamma(\bhalf+\i k+|l|)\,
                  \e^{\i l\vphi}{\cal P}^{-|l|}_{\i k-1/2}(\cosh\tau)\enspace,
\label{PSIkl}\\
  E_k&=&{\hbar^2\over2m}\bigg(k^2+{1\over4}\bigg)\enspace.
\label{Ekl}
\end{eqnarray}
For instance, we have the relation \cite{ABS}
\begin{eqnarray}
\CP_{\nu-1/2}^\mu(\cosh\tau)&=&\dfrac{1}{\Gamma(1-\mu)}
   2^{2\mu}(1-\e^{-2\tau})^{-\mu}\e^{-(\nu+1/2)\tau}
     \nonumber\\   & &\qquad\times
_2F_1\bigg(\half-\mu;\half+\nu-\mu;1-2\mu;1-\e^{-2\tau}\bigg)\enspace.
\end{eqnarray}

However, for the vector potential for an Aharonov--Bohm gauge field, we need 
another Ansatz. According to \cite{KURORU} I take for $\vec A=B\vec e_\vphi$ with 
$B=\hbox{const}$. Therefore I get for the classical Hamiltonian
\begin{equation}
 \CH=\dfrac{\hbar^2}{2m}\Bigg[p_\tau^2+\dfrac{1}{\sinh^2\tau}
      \bigg(p_\vphi-\dfrac{eB}{\hbar c}\bigg)^2\Bigg]\enspace,
\end{equation} 
and for the Lagrangian, respectively ($b=eB/\hbar c$) 
\begin{equation}
  \CL=\dfrac{m}2(\dot\tau^2+\sinh^2\tau\dot\vphi^2)+\dfrac{e}{c}\vec A\cdot
  \bigg(\begin{array}{c}\dot\tau\\ \dot\vphi\end{array}\bigg)
  =\dfrac{m}2(\dot\tau^2+\sinh^2\tau\dot\vphi^2)+\xi\dot\vphi\enspace.
\end{equation} 
Note that the vector potential in (\ref{Atau}) vanishes at $\tau=0$,
which means that we can take any constant for $A_\vphi$ depending on the
gauge, and the requirement that it is non-zero.
With the momentum operators $p_\tau=(\hbar/\i)(\partial_\tau+\coth\tau)$ and
$p_\vphi=(\hbar/\i)\partial_{\vphi}$ we get for the quantum Hamiltonian
(together with the quantum potential $\propto\hbar^2$)
\begin{equation}
 H=\dfrac{\hbar^2}{2m}\Bigg[p_\tau^2+\dfrac{1}{\sinh^2\tau}
      \bigg(p_\vphi-\dfrac{eB}{\hbar c}\bigg)^2\Bigg]
   +\dfrac{\hbar^2}{8m}\bigg(1-\dfrac{1}{\sinh^2\tau}\bigg)\enspace.
\end{equation} 
The angular variable $\vphi$ varies in the interval $[0,2\pi)$, and therefore
we usually assume $\vphi_j\in[0,2\pi),\forall_j$. However, the path can loop around 
the infinitesimal solenoid many times, which has the consequence that in our case
$\vphi_j\in\bbbr,\forall_j$. Therefore, the path integral, if calculated according to 
(\ref{Kpath-Bfeld}), gives only a partial propagator which belongs to a class of paths
topologically constraint by  $\vphi_j\in[0,2\pi),\forall_j$. For the total propagator,
we have to take into account all paths from all homotopically different classes. This 
can be done by considering the path integration over the angular variable $\vphi_j$ 
remaining in the physical space $\bbbm$ with $\Delta\vphi_j=\vphi_j-\vphi_{j-1}+2\pi \
n$ ($\vphi_j\in[0,2\pi),n\in\bbbz$), or alternatively switching to the covering space
$\bbbm^*$ with $\Delta\vphi_j=\vphi_j-\vphi_{j-1}$, where $\vphi_j\in\bbbr$.
I therefore incorporate the effect of the infinitesimal thin solenoid by a
$\delta$-function constraint in the path integral, with an additional integration
$\int\d\vphi$ \cite{BEIN}, therefore I get (expanding the $\delta$-function,
$\xi=e\Phi/2\pi\hbar c$ with $\Phi$ the magnetic flux.)
\begin{eqnarray}  & &\!\!\!\!\!\!\!\!
  K^{AB}(\tau'',\tau',\vphi'',\vphi';T)
    \nonumber\\   & &\!\!\!\!\!\!\!\!
  =\int_{\bbbr}\d\vphi\pathint{\tau}\sinh\tau\pathint{\vphi}
  \delta\bigg(\vphi-\int_0^T\dot\vphi\,\d t\bigg)
    \nonumber\\   & &\!\!\!\!\!\!\!\!\qquad\times
\exp\Bigg\{\ih\int_0^T\bigg[{m\over2}(\dot\tau^2+\sinh^2\tau\dot\vphi^2)+b\dot\vphi
  -\dfrac{\hbar^2}{8m}\bigg(1-\dfrac{1}{\sinh^2\tau}\bigg)\bigg]\d t\Bigg\}
    \nonumber\\   & &\!\!\!\!\!\!\!\!
  =\exp\bigg(-{\i\hbar T\over8m}\bigg)
  \int_{\bbbr}\d\vphi \int_{\bbbr}\dfrac{\d\lambda}{2\pi}\,\e^{\i\lambda\vphi}
 \lim_{N\to\infty}\bigg(\dfrac{m}{2\pi\i\hbar\epsilon}\bigg)^N\prod_{j=1}^{N-1}
  \int_0^\infty\sinh\tau_j\,\d\tau_j\int_0^{2\pi}\d\vphi_j\qquad\qquad
    \nonumber\\   & &\!\!\!\!\!\!\!\!\qquad\times
  \exp\left[\ih\sum_{j=1}^N\bigg({m\over2\epsilon}
  \Big(\Delta^2\tau_j+\widehat{\sinh^2\tau_j}\Delta^2\vphi_j\Big)
  +(\xi-\lambda)\Delta\vphi_j-{\epsilon\hbar^2\over8m\sinh^2\tau_j}
  \bigg)\right]\qquad
    \nonumber\\   & &\!\!\!\!\!\!\!\!
  =\int_{\bbbr}\d\vphi\int_{\bbbr}\dfrac{\d\lambda}{2\pi}\,\e^{\i\lambda\vphi}
  \sum_{l=-\infty}^\infty\e^{\i l(\vphi''-\vphi')}K_{\lambda+l-\xi}(\tau'',\tau';T)
  \enspace,
\label{Klblambda}
\end{eqnarray}
where 
\begin{eqnarray}  & &\!\!\!\!\!\!\!\!
  K_{\lambda+l-\xi}(\tau'',\tau';T) 
    \nonumber\\   & &
  =\e^{-\i\hbar T/8m}\pathint{\tau}\exp\Bigg[\ih\int_0^T\bigg({m\over2}\dot\tau^2-
 \dfrac{\hbar^2}{2m}\dfrac{(\lambda+l-\xi)^2-1/4}{\sinh^2\tau}\bigg)\d t\Bigg]\enspace.
\end{eqnarray}
Using Poisson's summation formula
\begin{equation}
  \sum_{l=-\infty}^\infty\e^{\i l\theta}=2\pi\sum_{k=-\infty}^\infty
  \delta(\theta+2\pi k)\enspace,
\end{equation}
I obtain (by changing the integration variable $\lambda\to\lambda+\xi-l$)
\begin{eqnarray}  & &\!\!\!\!\!\!\!\!
  K^{AB}(\tau'',\tau',\vphi'',\vphi';T)
    \nonumber\\   & &\!\!\!\!\!\!\!\!
  =\dfrac{1}{2\pi} \int_{\bbbr}\d\vphi \int_{\bbbr}\d\lambda\,\e^{\i\lambda\vphi}
  \sum_{l=-\infty}^\infty\e^{\i l(\vphi''-\vphi')}
  K_{\lambda+l-\xi}(\tau'',\tau';T)\qquad\qquad
    \nonumber\\   & &\!\!\!\!\!\!\!\!
  =\dfrac{1}{2\pi} \int_{\bbbr}\d\vphi \int_{\bbbr}\d\lambda\,
  \e^{\i l(\vphi''-\vphi'-\vphi)+\i(\lambda+\xi)\vphi} K_\lambda(\tau'',\tau';T)
     \nonumber\\   & &\!\!\!\!\!\!\!\!
 =\int_{\bbbr}\d\vphi \sum_{k=-\infty}^\infty\delta(\vphi''-\vphi'-\vphi+2\pi k)\,
  \e^{\i(\lambda+\xi)\vphi} \int_{\bbbr}\d\lambda\,K_\lambda(\tau'',\tau';T)\enspace.
\label{KABn}
\end{eqnarray}
$K_\lambda$ is now given by
\begin{eqnarray} 
  K_\lambda(\tau'',\tau';T)
                  &=&
  \e^{-\i\hbar T/8m}\pathint{\tau}\exp\Bigg[\ih\int_0^T\bigg({m\over2}\dot\tau^2-
 \dfrac{\hbar^2}{2m}\dfrac{\lambda^2-1/4}{\sinh^2\tau}\bigg)\d t\Bigg]
\label{pathtaulambda}
    \nonumber\\   &=&
  \int_0^\infty\d k\,\e^{-\i E_kT/\hbar}
  \Psi_{k,\lambda}(\tau'') \Psi_{k,\lambda}^*(\tau')\enspace.
\label{Ktaulambda}
\end{eqnarray}
The wave-functions and the energy spectrum are given by (\ref{PSIkl},\ref{Ekl}),
respectively, with $l\to\lambda$. Performing the $\vphi$-integration in 
(\ref{KABn}) yields
\begin{equation} 
  K^{AB}(\tau'',\tau',\vphi'',\vphi';T)=\sum_{n=-\infty}^\infty
  \e^{\i\xi(\vphi''-\vphi'+2\pi n)}\int_{\bbbr}\d\lambda\,
  \e^{\i\lambda(\vphi''-\vphi'+2\pi n)}K_\lambda(\tau'',\tau';T)\enspace,
\end{equation}
which displays the expansion into the winding numbers.
For $\xi=0$ the free Feynman kernel on $\Lambda$ is recovered.

If we want to study the effect of scattering by an Aharonov--Bohm solenoid we must 
consider interference terms according to
\begin{equation}
 I_{nl}=K^*_nK_l+K^*_lK_n\enspace.
\label{Inl}
\end{equation}
Unfortunately, a closed expression for the propagator (\ref{Ktaulambda}) does not 
exist. We can either analyze (\ref{Ktaulambda}) by means of an asymptotic 
expansion of the Legendre functions, i.e., $P_{\i p-1/2}^\mu(z)\propto (\Gamma(\i p)
/\Gamma(1/2+\i p-\mu))(2z)^{1/2-\i p}/\sqrt{\pi}+\hbox{c.c.}$, as $|z|\to\infty$, 
which yields very complicated and analytically intractable integrals over 
$\Gamma$-functions. Alternatively I can use the formula $\lim_{\nu\to\infty}\nu^\mu$
\linebreak
$\CP_\nu^{-\mu}(\cosh (z/\nu))=I_\mu(z)$ \cite{GRA} which corresponds to the flat 
space limit of the hyperbolic space with constant curvature $R$. Restricting 
therefore the evaluation of $I_{nl}$ to the flat space limit $R\to\infty$ I 
re-introduce the constant curvature $R$ into the path integral (\ref{pathtaulambda}) 
by means of $m\dot\tau^2\to mR^2\dot\tau^2=mr^2$, and $m\sinh^2\tau\to mR^2\sinh^2
\tau\to mR^2\tau^2=mr^2$ ($r=R\tau$ is the radial variable in Euclidean polar 
coordinates), as $R\to\infty$ \cite{IPSWa}. This gives for $K_\lambda$ in this limit 
the usual free Feynman kernel in polar coordinates in $\bbbr^2$ \cite{GRSh,PI}
\begin{equation}
K_\lambda(\tau'',\tau';T)\simeq K_\lambda(r'',r';T)
 =\dfrac{m}{2\pi\i\hbar T}\exp\bigg[\dfrac{\i m}{2\hbar T}({r'}^2+{r''}^2)\bigg]
 I_{|\lambda|}\bigg(\dfrac{mr'r''}{\i\hbar T}\bigg)\enspace.
\end{equation}
Following \cite{BEIN} we can now evaluate $I_{nl}$. By means of the asymptotic
formula ($|z|\to\infty,\Re(z)>0$)
\begin{equation}
  I_{\lambda}(z)\simeq\sqrt{1\over2\pi z}\,
  \exp\bigg(z-\dfrac{\lambda^2-1/4}{2z}\bigg)\enspace,
\end{equation}
and a Gaussian integration we get the asymptotic expansion
\begin{equation}
  \int_{-\infty}^\infty\d\lambda\,\e^{\i\lambda\Theta}I_\lambda(z)
  \simeq\exp\bigg(z+\dfrac{1}{8z}-\dfrac{z}{2}\Theta^2\bigg)\enspace.
\end{equation}
Hence I obtain for the partial propagator $K_n$ (with $z=mr'r''/\i\hbar T$, 
the condition $\Re(z)$ ignored, c.f.\ \cite{BEIN,GRSh,PI})
\begin{eqnarray}  & &\!\!\!\!\!\!\!\!
K_n(\tau'',\tau',\vphi'',\vphi';T)\simeq
\dfrac{m}{2\pi\i\hbar T}\exp\bigg[\dfrac{\i mR^2}{2\hbar T}(\tau''-\tau')^2
    \nonumber\\   & &\!\!\!\!\!\!\!\!\qquad\qquad
   +\dfrac{\i\hbar T}{8mR^2\tau'\tau''}+\i\xi(\vphi''-\vphi'+2\pi n)
   +\dfrac{\i mR^2\tau'\tau''}{2\hbar T}(\vphi''-\vphi+2\pi n)\bigg]\enspace.
\end{eqnarray}
Consequently, I get for the interference term
\begin{eqnarray} 
I_{nl}&\simeq&2\bigg(\dfrac{m}{2\pi\i\hbar T}\bigg)^2
    \nonumber\\   & & \!\!\!\!\!\!\!\!\!\!\!\!\!\!\!\!\times
\cos\Bigg[2\pi(l-n)\bigg(\xi+\dfrac{mR^2\tau'\tau''}{\hbar T}(\vphi''-\vphi'-\pi)\bigg)
+2\pi^2\dfrac{mR^2\tau'\tau''}{\hbar T}(l-n)(l+n+1)\Bigg]\enspace.\qquad
\label{Inl2}
\end{eqnarray}
The principal feature of this result consists that the interference patterns
does not depend only on the initial $(\tau',\vphi')$ and final points
$(\tau'',\vphi'')$, but on the homotopy class numbers $n$ and $m$ as well which 
describe the windings around the infinitesimal thin solenoid. This flux dependent 
shift is a proper Aharonov--Bohm effect. The interference term vanishes for $n=l$.

The maximum contribution to the Aharonov--Bohm effect on the (hyperbolic) plane is 
observed for the smallest non-vanishing value $|n-l|=1>0$. Therefore, the maximum 
effect is observed for the interference of the winding number $l=0$ and $n=-1$, or 
vice versa, yielding the interference term
\begin{equation}
I_{0,-1}=2\bigg(\dfrac{m}{2\pi\i\hbar T}\bigg)^2\cos(2\pi\xi)\enspace.
\end{equation}
This is the standard result, e.g.~\cite{FH} and \cite{BEIN} and references therein.

\section{Higgs-Oscillator and Kepler--Coulomb Potential}
\message{Higgs-Oscillator and Kepler--Coulomb Potential}
Obviously, we can incorporate potential terms in the radial path integration
$\tau$, e.g., we can include the Higgs-oscillator potential \cite{GROPOc,HIGGS}
\begin{equation}
  V_{(\Higgs)}(\vec u)={m\over2}\omega^2R^2{u_1^2+u_2^2\over u_0^2}
 ={m\over2}\omega^2R^2\tanh^2\tau\enspace,
\end{equation}
which is the analogue of the harmonic oscillator in a space of constant 
curvature, or the Kepler--Coulomb potential \cite{BIJb,GROe,GROPOc}, respectively
\begin{equation}
  V_{(\Coulomb)}(\vec u)=-{\alpha\over R}\left({u_0\over\sqrt{u_1^2+u_2^2}}-1\right)
  =-{\alpha\over R}(\coth\tau-1)\enspace.
\end{equation}
For clarity, I have included the dependence on the constant curvature $R$
explicitly. In these cases, the result (\ref{Klblambda}) is more appropriate.
The combined $\d\vphi\,\d\lambda$-integration yields $\lambda=0$, and the total
propagator becomes
\begin{equation}
  K^{AB}(\tau'',\tau',\vphi'',\vphi';T)=\sum_{l=-\infty}^\infty
  e^{\i l(\vphi''-\vphi')}\,K_{|l-\xi|}(\tau'',\tau';T)\enspace,
\end{equation}
and the effect of the solenoid exhibits in a modification of the angular momentum 
dependence of $K_{|l-\xi|}$. This feature, however, modifies the number of bound 
states of the system with respect to the quantum number $l$. For instance, for the 
Higgs-oscillator case this gives ($\nu^2=m^2\omega^2R^4/\hbar^2+1/4$)
\begin{eqnarray}       & &\!\!\!\!\!\!\!\!\!\!\!\!\!\!
  \Psi_{nl}^{(\Higgs)}(\tau,\vphi;R)=(2\pi\sinh\tau)^{-1/2}
  S_n^{(\nu)}(\tau;R)\,\e^{\i l\vphi}\enspace,
                  \\   & &\!\!\!\!\!\!\!\!\!\!\!\!\!\!
  S_n^{(\nu)}(\tau;R)={1\over\Gamma(|l-\xi|+1)}
  \bigg[{2(\nu-|l-\xi|-2n-1)\Gamma(n+|l-\xi|+1)\Gamma(\nu-|l-\xi|)\over
         R^2\Gamma(\nu-|l-\xi|-n)n!}\bigg]^{1/2}
         \nonumber\\   & &\!\!\!\!\!\!\!\!\!\!\!\!\!\!
  \qquad\qquad\times
  (\sinh\tau)^{|l-\xi|+1/2}(\cosh\tau)^{n+1/2-\nu}
  {_2}F_1(-|l-\xi|,\nu-n;1+|l-\xi|;\tanh^2\tau)\enspace,
\end{eqnarray}
with the discrete spectrum given by
\begin{equation}
  E_n^{(\Higgs)}=-\dfrac{\hbar^2}{2mR^2}\bigg[(2n+|l-\xi|-\nu+1)^2-\viert\bigg]
      +{m\over2}\omega^2R^2\enspace.
\end{equation}
Only a finite number exist with $N_{\max}=[\nu-|l-\xi|-1]\geq0$ ($[x]$ denotes the 
integer value of $x\in\bbbr$). The continuous wave-functions have the form
\begin{eqnarray}       & &\!\!\!\!\!\!\!\!\!\!\!\!\!\!
  \Psi_{kl}^{(\Higgs)}(\tau,\vphi;R)=(2\pi\sinh\tau)^{-1/2}
  S_k^{(\nu)}(\tau;R)\,\e^{\i l\vphi}\enspace,
                  \\   & &\!\!\!\!\!\!\!\!\!\!\!\!\!\!
  S_k^{(\nu)}(\tau;R)={1\over\Gamma(|l-\xi|+1)}\sqrt{k\sinh\pi k\over2\pi^2R^2}\,
  \Gamma\bigg({\nu-|l-\xi|+1-\i k\over2}\bigg)
  \Gamma\bigg({|l-\xi|-\nu+1-\i k\over2}\bigg)
         \nonumber\\   & &\!\!\!\!\!\!\!\!\!\!\!\!\!\!\qquad\times
  (\tanh\tau)^{|l-\xi|+1/2}(\cosh\tau)^{\i k}
         \nonumber\\   & &\!\!\!\!\!\!\!\!\!\!\!\!\!\!\qquad\times
  {_2}F_1\bigg({\nu+|l-\xi|+1-\i k\over2},{|l-\xi|-\nu+1-\i k\over2}
               ;1+|l-\xi|;\tanh^2\tau\bigg)\enspace,\qquad
\end{eqnarray}
with the continuous energy-spectrum given by
\begin{equation}
  E_p^{(\Higgs)}={\hbar^2\over2mR^2}\bigg(k^2+\viert\bigg)+{m\over2}\omega^2R^2
  \enspace.
\end{equation}
In the case of the Kepler--Coulomb problem on $\Lambda$ we obtain for the discrete 
energy spectrum ($\tilde N=N+|l-\xi|+\half$, ($N=0,1,2,\dots,N_{\max}=[\sqrt{R/a}-
|l-\xi|-\half], a=\hbar^2/m\alpha$ is the Bohr radius)
\begin{equation}
  E_N^{(\Coulomb)}={\alpha\over R}-\hbar^2{\tilde N^2-\viert\over2mR^2}
  -{m\alpha^2\over2\hbar^2\tilde N^2}\enspace.
\end{equation}
The wave-functions I do not state, c.f.\ \cite{GROPOc}, and the continuous
states are modified by their angular momentum dependence, i.e., $l\to l-\xi$. 
However, the effect of the Aharonov--Bohm field is not only restricted to a 
modification of the discrete spectrum, but the effect on the scattering states 
happens through an interference term $I_{nl}$ similarly to (\ref{Inl}), for the 
Coulomb potential and the Higgs-oscillator as well. Again, a closed expression 
for the radial propagator does not exist and we are restricted to the investigation 
of the limiting case along the lines following (\ref{Inl}). 
This I do not repeat once more.

\newpage\vspace*{-1.0cm}\noindent%
\section{Summary}
\message{Summary}
I therefore have shown the admissibility of path integration of the Aharonov--Bohm 
effect on the hyperbolic plane. It can be studied in a straightforward manner 
yielding analogous results in comparison to the flat space case. For scattering 
states we find interference, due to the modification of the angular momentum
dependence according to $l\to l-\xi$, giving a $\cos$-like pattern in terms of the 
strength of the vector-potential, for the free motion, the Kepler--Coulomb problem, 
and the Higgs-oscillator (which is absent in the flat space case); the  bound state 
wave-function and the corresponding energy levels are modified in their
angular momentum dependence $l\to l-\xi$ as well, together including an alteration of 
the number of bound states. We found the usual expansion of the total propagator in 
terms of an expansion into the winding number $n$ of the homotopy class of paths. 
All these features are well-known form the corresponding flat-space cases. The 
complicated interference expression (\ref{Inl}) could not be evaluated due the 
non-constant curvature features of the hyperbolic plane. This would involve an 
analytical intractable integration over Legendre functions with respect to the order. 
However, the investigation of the flat space-limit gave the well-known result. 
Therefore the effect of an Aharonov--Bohm gauge field on the hyperbolic plane, i.e., 
scattering on leaky tori,  exhibits the same features as in the flat space case of 
$\bbbr^2$.


\bigskip\bigskip 
\vbox{\centerline{\ }
\centerline{\quad\epsfig{file=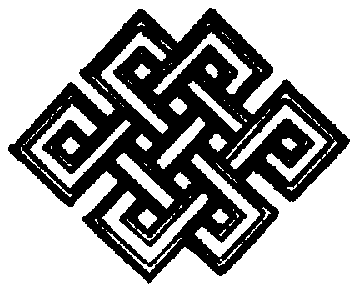,width=4cm,angle=90}}}

\end{document}